\documentclass[prb]{revtex4}% Physical Review B

\usepackage{graphicx}% Include figure files
\usepackage{dcolumn}% Align table columns on decimal point
\usepackage{amsmath,amssymb}
\usepackage{bm}% bold math

\begin{document}

%\preprint{APS/123-QED}

\title{A theoretical investigation into the microwave spectroscopy of a phosphorus-donor charge-qubit in silicon: Coherent control in the Si:P quantum computer architecture}

\author{C.J.  Wellard, L.C.L Hollenberg}
%\email{cjw@physics.unimelb.edu.au}
 \affiliation{Centre for Quantum Computer Technology, School of Physics, University of Melbourne, Australia.}
\author{S.  Das Sarma}
\affiliation{Condensed Matter Theory Centre, Department of Physics, University of Maryland.}

\date{\today}% It is always \today, today,

\begin{abstract}
We present a theoretical analysis of a microwave spectroscopy experiment on a charge qubit defined by a P$_2^+$ donor pair in silicon, for which we calculate Hamiltonian parameters using the effective-mass theory of shallow donors.  We solve the master equation of the driven system in a dissipative environment to predict experimental outcomes.  We describe how to calculate physical parameters of the system from such experimental results, including the dephasing time, $T_2$, and the ratio of the resonant Rabi frequency to the relaxation rate.  Finally we calculate probability distributions for experimentally relevant system parameters for a particular fabrication regime.

\end{abstract}

\pacs{Valid PACS appear here}% PACS, the Physics and Astronomy

\maketitle

\section{introduction}
The strong motivation supplied by the possibility of quantum information processing has recently fuelled rapid progress in the experimental control of mesoscopic quantum systems.  Of particular interest in the solid-state are superconducting devices, and coupled quantum dots, which show promise as candidates for qubits in a scalable quantum device.  In order to realise this potential it is necessary that these systems be coherently controlled with extraordinary precision, and much progress has been made toward this goal in superconducting systems \cite{Nakamura99,Nakamura01,Vion03}, as well as  coupled quantum dots in gallium arsenide \cite{Hayashi03}, and phosphorus doped silicon \cite{Gorman05}.

In terms of the experimental demonstrations of coherent control one can take two approaches, the observation of Ramsey fringes in a pulsed experiment \cite{Nakamura99,Hayashi03,Vion03,Gorman05}, where the coherent oscillations are between non-diagonal eigenstates of the system Hamiltonian.  The system must be manipulated on a timescale that is fast compared to the period of the oscillations, which is inversely proportional energy difference of the eigenstates.  Alternatively one could  observe the Rabi oscillations in a driven system \cite{Nakamura01,Vion03}, where the frequency of the oscillations is proportional to the intensity of the driving field.

 In many cases a precursor to these achievements is the somewhat simpler procedure of performing a spectroscopic measurement of the device \cite{Nakamura97,Oosterkamp98,Petta04}.  This is achieved by driving the system of interest at particular frequency, and varying parameters to bring the system into resonance.  When in resonance, the system will undergo Rabi oscillations, however a spectroscopic experiment of this nature does not resolve these oscillations, instead measurements are made on a timescale that is greater than the Rabi frequency, giving a time averaged result.  Useful results can be obtained even in the regime in which the frequency of the oscillation is slow compared with relevant decoherence times.  Clearly a measurement of this type places far fewer demands on the system than if the fringes are resolved, however much useful information can be gained about the system, such as the dephasing time $T_2$, some of the system energy scales, and the ratio of the resonant Rabi frequency to the relaxation rate.

One of the most promising quantum computer architectures, particularly from
the all-important scalability perspective, is the phosphorus-doped silicon
(Si:P) system where the shallow P-bound electronic donor states are
electrically manipulated by external gates for the quantum control of the
qubits.  In spite of impressive recent growth, fabrication, and theoretical
developments in the Si:P quantum computer architecture, the key ingredient
of an experimental demonstration of coherent quantum dynamics at the single
qubit level is still lacking.  In this paper we consider a specific
experimental set-up, namely the externally driven microwave spectroscopy on
a P$_2^+$ charge qubit \cite{Hollenberg04} (i.e. the singly ionised coupled P-P effective
'molecular' system in Si with two nearby ion implanted P atoms sharing one
electron), giving a detailed quantitative theoretical analysis to establish
the feasibility of this coherent single qubit control experiment in
currently available Si:P devices.  We believe that the qubit coherent
control experiment proposed and analysed in this work may very well be the
simplest single qubit measurement one could envision in the Si:P quantum
computer architecture, and as such, our detailed quantitative analysis
should motivate serious experimental efforts using microwave spectroscopy.
The experiment being analysed in this work is a necessary prerequisite for
further experimental progress in coherent qubit control and manipulation in
the Si:P quantum computer architecture.  Our detailed theory provides the
quantitative constraints on the experimental system parameters required for
developing the coherent control of the Si:P system, and in addition,
establishes how one can extract important decoherence parameters (e.g. $T_2$)
from the experimental data.

The energetics of this system, including its interaction with phonon modes in the lattice\cite{Hu05}, and with a DC electric field \cite{Koiller05}, have been studied by previous authors, while a theoretical investigation of the microwave spectroscopy of a coupled quantum dot in  a GaAs system has been carried out by Barrett {\it et al} \cite{Barrett05}, and much work has been done on the theory of the dynamics of driven systems coupled to dissipative environments has been well studied in the context of flux qubits, \cite{Smirnov03A,Smirnov03B}.

In this article we begin by reviewing the $P_2^+$ in silicon system in the presence of an electric field, and outline the calculation of the Hamiltonian matrix elements for use in the calculation of the system dynamics.  In sec~\ref{sec:master} we derive a master equation for the driven system, in the presence of three decoherence channels, following the the approach of Barrett \cite{Barrett05}.  This master equation is analytically solved for the steady-state solutions, valid on a time-scale that is long compared to the relaxation time.  To probe the dynamics on shorter time-scales requires, in general, numerical integration of the master equation.  In sec~\ref{sec:results} we show how it is possible to use the output of the measurement device to obtain the energy scales of the system, and to calculate the dephasing time $T_2$, while in sec.~\ref{sec:coherent} we briefly discuss possibilities for the direct observation of coherent oscillations in this system.  Finally, in sec~\ref{sec:stats} we discuss the statistical nature of the device fabrication, we calculate probability distributions for experimentally relevant parameters, and discuss likely experimental outcomes.

\section{A two-donor charge qubit in silicon}
\label{sec:p2+}
We consider a system defined by the electronic state of a pair of phosphorus donors in silicon, which has been singly ionised such that the system has a single net positive charge \cite{Hollenberg04}.  The donor nuclei are separated by a vector ${\bf R} $, and the in the absence of externally applied fields, the  system Hamiltonian is
\begin{equation}
 H_0 = H_{Si} +  V_d({\bf r}-{\bf R}/2) +  V_d({\bf r}+{\bf R}/2).  
\end{equation}
Here $H_{Si}$ is the Hamiltonian of an electron in the pure silicon lattice, which includes both a  kinetic term and the effective potential due to the silicon lattice.  Solutions of this silicon Hamiltonian are the Bloch functions of the pure silicon crystal.  The $V_d({\bf r})$ terms give the impurity potential of the ionic donor cores, which we will treat as Coulombic, $V_d({\bf r}) = -2 /\kappa r$, with $\kappa = 11.9$, the dielectric constant of silicon.  

 The electronic ground state for a single donor, centred at a position ${\bf R}/2$ is, in the effective-mass approximation, given by the Kohn-Luttinger wavefunction \cite{Kohn57}
\begin{equation}
\psi({\bf r}) = \sum_\mu F_\mu({\bf r}-{\bf R}/2) u_{{\bf k}_\mu}({\bf r}) e^{i {\bf k}_\mu.({\bf r} -{\bf R}/2)}.
\end{equation}
In this equation the sum is over the 6 degenerate minima of the silicon conduction band, ${\bf k}_\mu$ denoting the reciprocal lattice vector at the minima $\mu$.  The function $u_{{\bf k}_\mu}({\bf r})$ is the periodic part of the conduction band Bloch function, and $F_{\mu}({\bf r})$ is a non-isotropic, hydrogen-like envelope function, which for the $\pm z$ minima takes the form
\begin{equation}
F_{\pm z} = \frac{e^{-\sqrt{ z^2/ a_\parallel^2 + (x^2+y^2)/a_\perp^2 }}}{\sqrt{ 6 \pi a_\perp^2 a_\parallel}},
\end{equation}
where $a_\parallel,a_\perp$ are the non-isotropic effective masses associated with the conduction band minima.

\begin{figure}
\rotatebox{0}{\resizebox{7cm}{!}{\includegraphics{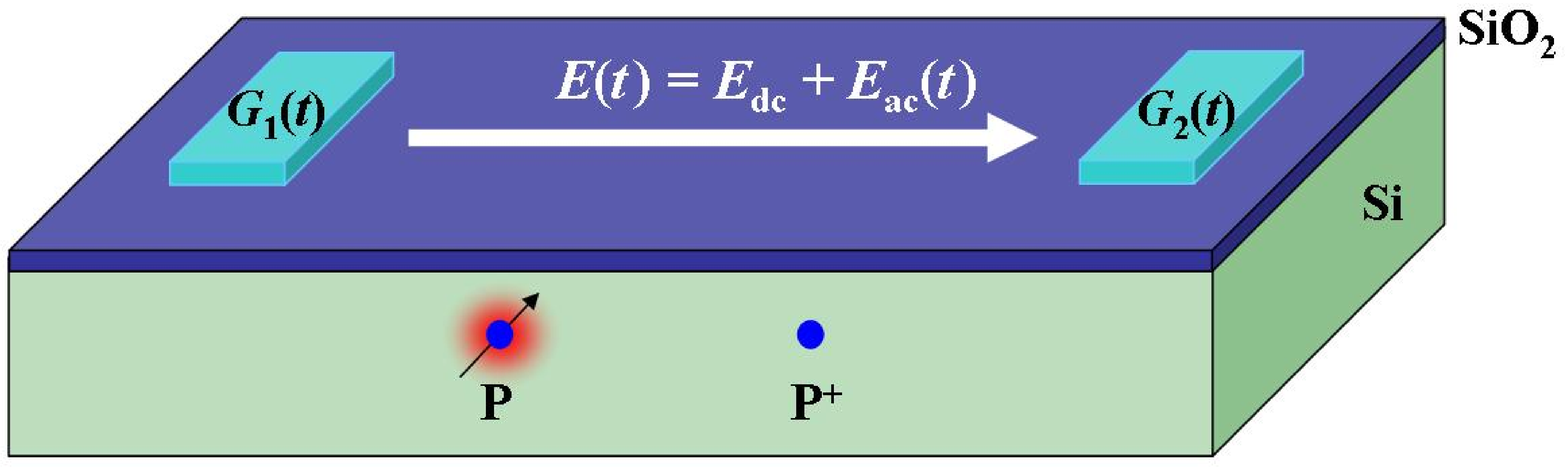}}}
\rotatebox{0}{\resizebox{7cm}{!}{\includegraphics{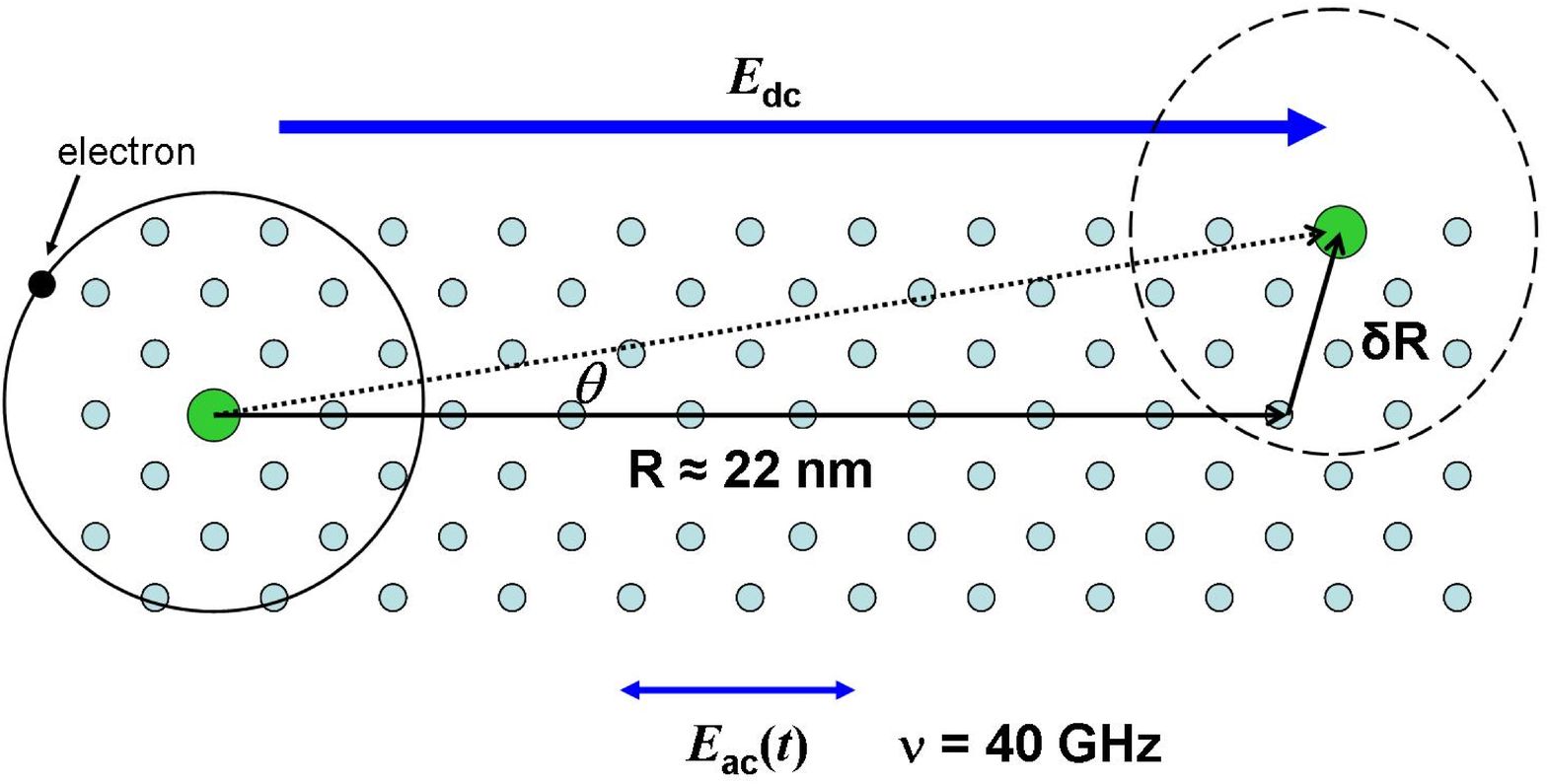}}}
\caption{(Colour online) On the left is a schematic of the experimental setup showing the location of the bias gates relative to the phosphorus donors. On the right is a diagram showing orientation of the two donors, and the direction of the applied fields, relative to the silicon crystal lattice.}\label{fig:archi}
\end{figure}

\begin{figure}
\rotatebox{0}{\resizebox{9cm}{!}{\includegraphics{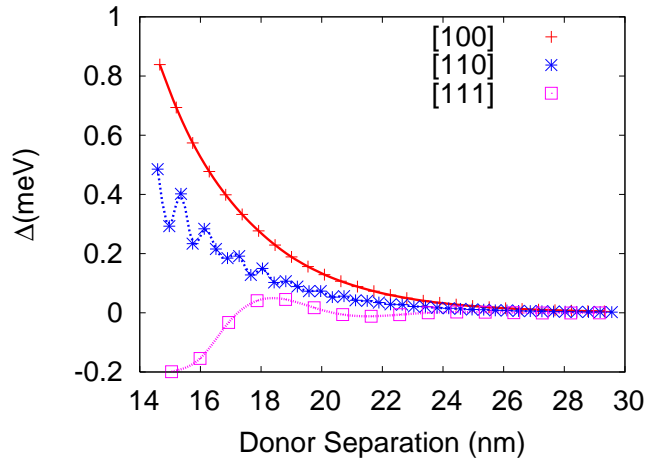}}}
\caption{(Colour online) A plot of the symmetric anti-symmetric energy gap $\Delta_{\rm s-as}$ as a function of donor separation along three high symmetry crystallographic axes.  The points denote fcc lattice sites, the lines are a guide to the eye.}\label{fig:gap}
\end{figure}

If $R$ is sufficiently large, the low-lying energy states of the two-centre system,  are approximated by even and odd superpositions of the single donor ground-states, $ \Psi_\pm({\bf r})  = 1/\sqrt{2(1+S^2)}(\psi({\bf r}-{\bf R}/2)  \pm \psi({\bf r}+{\bf R}/2) )$, where $S = \int \psi^*({\bf r}-{\bf R}/2) \psi({\bf r}+{\bf R}/2) d{\bf r}$.  
Interference between Bloch functions at the different conduction-band minima leads to a non-monotonic dependence of the symmetric-antisymmetric energy gap, $\Delta_{\rm s-as}$, on the magnitude of the donor separation, as well as on their orientation with respect to the silicon substrate \cite{Hu05}.  This is illustrated in Fig.~\ref{fig:gap}, where we plot $\Delta_{\rm s-as}$ as a function of donor separation, for donors separated along three high symmetry crystallographic axes.  Note that for certain separations, the ground state is actually the antisymmetric state.

In the basis of the single donor ground states , we can write
\begin{equation}
H_0 = h_x \sigma^x + h_z \sigma^z,
\end{equation}
where the coefficients $H_\alpha$ are determined by numerical integration for various vectors ${\bf R}$.  Note we have ignored the constant term, and defined our axes such that the $\sigma^y$ component is zero.  In  practice we expect that the two donors are identical, in which case $H_z=0$, however we retain this term in what follows for generality.

 The addition of a uniform electric field in a direction along a [100] axis, gives rise to a potential that can be written as
\begin{equation}
V = E (\xi_x \sigma^x + \xi_z \sigma^z),
\end{equation}
where $E$ is the strength of the applied field, and the coefficients $\xi_x,xi_z$ are determined by numerical integration.  In the case of radially symmetric single donor wavefunctions, the $\xi_x$ term is identically zero, however this symmetry is broken by the silicon lattice.  In practice, we find, $\xi_z>>\xi_x$.  The magnitude this difference is dependent on ${\bf R}$, but for $R\approx20$nm, it is typically a factor of between 100-1000.

To begin with, we consider the effect of a DC field
\begin{eqnarray}
H &=& H_0 +   V_{\rm dc}  \nonumber\\
 &=& (h_x + E_{\rm dc} \xi_x )\sigma^x + (h_z+E_{\rm dc} \xi_z) \sigma^z \nonumber\\
&=& \Theta \sigma^z {\rm e}^{i \phi \sigma^y}.
\end{eqnarray}
Here $\Theta = \sqrt{(h_x + E_{\rm dc} \xi_x )^2 + (h_z + E_{\rm dc} \xi_z )^2}$, and $\phi = {\rm arctan}\left[(h_x + E_{\rm dc} \xi_x )/(h_z+ E_{\rm dc} \xi_z )\right]$.  Thus, the above Hamiltonian can be diagonalised via the transformation:
\begin{equation}
H' = U_1 H U_1^\dag = \Theta \sigma^z,
\end{equation}
with $U_1 = \exp(i \phi /2  \sigma^y )$.  The eigenvalues are given by $\lambda_{\pm} = \pm \Theta$, and the eigenstates, in the original charge localised basis, by:
\begin{eqnarray}
|+\rangle &=& \cos[\phi/2]|L\rangle -\sin[\phi/2] |R\rangle, \nonumber\\
|-\rangle &=& \sin[\phi/2] |L\rangle + \cos[\phi/2] |R\rangle.
\end{eqnarray}

Thus by varying the strength of the field we can alter the energy splitting $\Delta = 2 \Theta$, between the two lowest energy states, as well as the degree of localisation of the eigenstates $\chi_\pm = |<L|\pm>|^2 = 1/2(1 \pm \cos(\phi))$.

We now consider the additional of an AC field of angular frequency $\omega$ applied to  the uniform DC field.  The Hamiltonian of this driving field is 
\begin{equation}
V_{\rm ac} = E_{\rm ac} \cos(\omega t) (\xi_x \sigma^x + \xi_z \sigma^z),
\end{equation}
which transforms to the energy eigenbasis of the time independent system as 
\begin{eqnarray}
V_{\rm ac}' &=& U_1 V_{\rm ac} U_1 \nonumber\\
&=& E_{\rm ac} \cos(\omega t) (\sigma^x \gamma_x  + \sigma^z \gamma_z ),
\end{eqnarray}
with $\gamma_x = \xi_x \cos(\phi) -\xi_z \sin(\phi), \gamma_z = \xi_z \cos(\phi) + \xi_x  \sin(\phi)$.
We now transform into an interaction picture, rotating at the same frequency as the AC field, the transformation is defined by $U_2(t) = {\rm e}^{i \omega/2 t \sigma^z}$.  Evolution in this frame is generated by the Hamiltonian
\begin{equation}
{\tilde H}' = (\Theta-\omega/2) \sigma^z +  E_{\rm ac} \cos(\omega t) ( \sigma^x \gamma_x {\rm e}^{-i \omega t \sigma^z} + \sigma^z \gamma_z ).
\end{equation}
It is, at this point that it becomes convenient to make a rotating wave approximation (RWA), in which we ignore terms that are rotating with frequency $\omega$.  This approximation is valid if $\omega >>  E_{\rm ac} \gamma_x , E_{\rm ac} \gamma_z$, and reduces the above Hamiltonian to 
\begin{equation}
{\tilde H}'=(\Theta-\omega/2) \sigma^z + E_{\rm ac} \frac{ \gamma_x }{2} \sigma_x.
\end{equation}

\section{Construction of the master equation}
\label{sec:master}
To include the effects of a dissipative environment on the dynamics of the system, it is necessary to construct a master equation.  We consider the effects of three channels of decoherence, namely;  dephasing, relaxation and excitation, with different rates $\Gamma_z,\Gamma_-,\Gamma_+$ respectively.  We expect these channels to be sufficient to describe most physical noise sources \cite{Gardiner91,Carmichael93} including interaction with the measurement device \cite{Barrett05,Schoelkopf02}.  Generally we would expect, $\Gamma_z $ to be the dominant term, with  the relaxation term $\Gamma_- < \Gamma_z$.  The excitation term is included for generality, and allows the description of the effects of classical, or high temperature noise sources, we expect $\Gamma_+ \leq \Gamma_-$.

These decoherence channels are included in the master equation, in the energy eigenbasis, as Linbladian terms  ${\cal L}\{ L_i,\rho \} =  L_i \rho L_i^\dag  -1/2 (L_i^\dag L_i \rho + \rho L_i^\dag L_i ) $.  thus, in the rotating frame, the master equation is given by
\begin{eqnarray}
\dot{{\tilde \rho}}'(t) &=& - i \left[(\Theta-\omega/2) \sigma^z +E_{\rm ac} \frac{\gamma_x }{2} \sigma_x,{\tilde \rho}'(t)\right] \nonumber\\
&+&\Gamma_z{\cal L}\{ \sigma^z, {\tilde \rho}'(t)\}  + \Gamma_-{\cal L}\{\sigma^-, {\tilde \rho}'(t)\} + \Gamma_+{\cal L}\{ \sigma^+, {\tilde \rho}'(t)\},
\end{eqnarray}

The steady state solution of this master equation can be found, which gives the asymptotic form for the components of the polarisation vector, defined by $\rho = 1/2(\mathbb{I} + X \sigma_x + Y \sigma_y + Z \sigma_z)$, in the rotating frame: 
\begin{eqnarray}
{\tilde X}'(\infty) &=& \frac{-8 \eta \gamma (\Gamma_+-\Gamma_-)}{16 (\Gamma_++\Gamma_-)\eta^2 +(2 _(E_{\rm ac} \gamma_x)^2+ (\Gamma_++\Gamma_-)(\Gamma_++\Gamma_-+4\Gamma_z))(\Gamma_++\Gamma_-+4\Gamma_z) } \nonumber\\
{\tilde Y}'(\infty) &=& \frac{2 \eta \gamma (\Gamma_+-\Gamma_-)(\Gamma_++\Gamma_-+4\Gamma_z)}{16 (\Gamma_++\Gamma_-)\eta^2 +(2 (E_{\rm ac} \gamma_x)^2+ (\Gamma_++\Gamma_-)(\Gamma_++\Gamma_-+4\Gamma_z))(\Gamma_++\Gamma_-+4\Gamma_z) } \nonumber\\
{\tilde Z}'(\infty) &=& (\Gamma_+-\Gamma_-)\frac{16 \eta^2 + (\Gamma_++\Gamma_-+4\Gamma_z)^2}{16 \eta^2 (\Gamma_++\Gamma_-) +(2 (E_{\rm ac} \gamma_x)^2+ (\Gamma_++\Gamma_-)(\Gamma_++\Gamma_-+4\Gamma_z))(\Gamma_++\Gamma_-+4\Gamma_z) } .
\end{eqnarray}
where $\eta = \Theta-\omega/2$ is the detuning of the microwave field.

We wish to calculate the expected output from a detector which measures charge localisation, a $z$-measurement in the original basis, and therefore are interested in calculating $Z$-component of the polarisation vector in the charge localised basis and the laboratory frame, the asymptotic form of which is given by
\begin{equation}
Z(t)= ({\tilde Z}' \cos(\phi) - (\cos(\omega t) {\tilde X}' + \sin(\omega t) {\tilde Y}') \sin(\phi)).
\end{equation}

In the case that the measurement time of the detector is long compared to both the relaxation timescale, $t_{\rm det} > 1/\Gamma_{-}$, and the timescale of these oscillations $t_{det} > 1/\Omega$, then the system reaches the steady state equilibrium over the time of the measurement, and the the rapid oscillations of the rotating frame will average out.  In this case the probability of measuring the electron to be localised at the position of the left donor, is given by ${\rm P}_{\rm L} = (1+\cos(\phi){\tilde Z}'(\infty))/2$.

\section{Application}
\label{sec:results}

\begin{figure}
\rotatebox{0}{\resizebox{9cm}{!}{\includegraphics{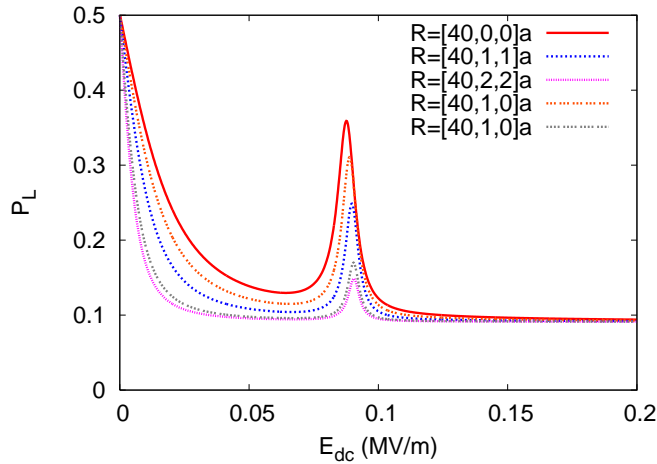}}}
\caption{(Colour online) Donor spectroscopy for several donor separations with a small deviation from the ideal separation of $R = [40a,0,0]$, where $a=0.543$nm is the lattice constant of silicon.  The data shows that the peak heights and positions are not strongly dependant on these positional variations, however the width of the peak changes significantly.  All data were calculated with a $40$GHz driving field of amplitude ${\rm E_{ac}} = 10^{-2} {\rm MV/m} $, and decoherence rates $\Gamma_z = 3{\rm GHz}, \Gamma_- = 1{\rm GHz}$ and $\Gamma_+ = 100{\rm MHz}$.}\label{fig:dev}
\end{figure}

We now turn our attention to analysis of simulated experimental results.  In practice the P-P$^+$ charge qubit is, in the first instance, fabricated via ion implantation \cite{Jamieson05}, whereby 14keV phosphorus ions are implanted into a silicon crystal.  Such a process is imprecise and there will be significant uncertainty in the position of the two phosphorus donors.  As we have previously discussed, this will be manifest in an uncertainty in both the bare Hamiltonian of the two donor system, as well as in the potential arising from the applied electric fields.  To illustrate how this uncertainty will affect experimental results, we have calculated simulated experimental traces, $P_L$ as a function of applied DC field $E_{\rm dc}$, for donor pairs with with slightly varying separations, using  a $\nu = 40{\rm GHz}$ driving field of amplitude ${\rm E}_{\rm ac} = 10^{-2} {\rm MVm}^{-1}$.  The results are shown in Fig.~\ref{fig:dev}, and show that although a slight variation of the donor separation has little effect on the position of the resonance peak, both the width and height of the peak are strongly affected.

\begin{figure}
\rotatebox{0}{\resizebox{9cm}{!}{\includegraphics{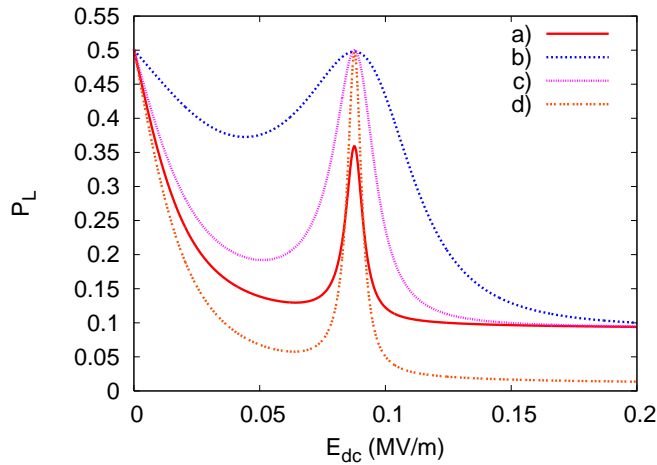}}}
\caption{(Colour online) An illustration of the effects of various decoherence channels on the spectroscopic signal height.  These channels affect both the peak height and width as described in the text.  The lengths are in units of $a=0.543$nm, the lattice constant of silicon, all data in this plot was taken for a donor separation $R=[40a,0,0]$, where $a=0.543$nm is the lattice constant of silicon. The frequency of the driving field was 40GHz and its amplitude  $E_{ac} = 0.01 {\rm MVm^{-1}}$.  The decoherence rates are a) $\Gamma_Z = 3$GHz,\ $\Gamma_-=1$GHz,\ $\Gamma_+=100$MHz,\ b)$\Gamma_Z = 3$GHz,\ $\Gamma_-=10$MHz,\ $\Gamma_+=1$MHz,\ c)$\Gamma_Z = 300$MHz,\ $\Gamma_-=10$MHz,\ $\Gamma_+=1$MHz,\ c)$\Gamma_Z = 300$MHz,\ $\Gamma_-=100$MHz,\ $\Gamma_+=1$MHz.}\label{fig:rwa2}
\end{figure}

In addition to the system Hamiltonian parameters, an experimental trace will be effected by the strength of the decoherence channels, as shown in Fig.~\ref{fig:rwa2}.  Here we see that both the height, and the width of the peaks are dependant of the rates of decoherence, whilst the position of the peak, is determined by the eigen energies of the system.  It is, therefore, reasonable to expect that useful information about the system can be obtained from the analysis of this spectroscopic data.

Given an experimental trace, we would like to determine useful spectroscopic information about the qubit, in particular we would like to determine decoherence times, as well as the energy scales of the system.   Observation of the height of the baseline, taken at a point where this baseline is flat, allows the evaluation of the ratio $\Gamma_+/\Gamma_-$.  In the absence of microwaves, the signal will be $P_L = 1/2(1+(\Gamma_+-\Gamma_-) \cos(\phi)/(\Gamma_++\Gamma_-))$, which allows the evaluation of $\cos(\phi)$ over the range of $E_{\rm dc}$.

In general, the strength of the applied electrostatic fields at the qubit, both AC and DC, may not be known for a given control gate bias.  In what follows we will assume that both these quantities are linear functions of their respective voltage biases, $E_{\rm dc} = \alpha G_{\rm dc},  E_{\rm ac} = \beta G_{\rm ac}$ where the $G_{\rm dc},G_{\rm ac}$ are known experimental control parameters.   

By obtaining several traces taken at different driving frequencies, and observing the positions of the resonance peaks $E^*_{dc}$, it is possible to obtain values for $\Theta$ as a function of $G_{\rm dc}$.  Combined with the knowledge of $\cos{\phi}$, this allows us to obtain values for $\alpha \xi_z$ and $h_x$.  

Additional information can be gained from measuring the width of the resonance peak, the half-width at half maximum, measured from the flat baseline value, is given by the expression

\begin{equation}
\Delta G_{dc} = \frac{1}{2 \alpha \xi_z} \sqrt{\frac{(2(E_{\rm dc} \gamma_x)^2+(\Gamma_++\Gamma_-)(\Gamma_++\Gamma_-+4\Gamma_z))(\Gamma_++\Gamma_-+4\Gamma_z)}{\Gamma_++\Gamma_-}}\label{eq:width}
\end{equation}.

Here we have assumed that $\cos(\phi)=1$ at the position of the resonance, which is a good approximation in all the traces shown in this article.  If this is not the case, a slightly more complicated expression can be derived for Eq~.\ref{eq:width}.  If traces can be evaluated for different values of the microwave power, this expression gives information about the decoherence rates, in particular extrapolation to zero microwave power yields $\Delta G_{\rm dc} = 1/(2 \alpha \xi_z) (\Gamma_++\Gamma_-+4\Gamma_z)$.  Given that $\alpha \xi_z$ has already been determined, this allows direct evaluation of the dephasing time $T_2 = 4/(\Gamma_++\Gamma_-+4 \Gamma_z)$.
 An expression for the height of the resonance peak,
\begin{equation}
h \approx \frac{1}{2}\left(1 + \frac{ (\Gamma_+-\Gamma_-)(\Gamma_++\Gamma_-+4\Gamma_z) }{2 (E_{\rm ac} \gamma_x)^2 + (\Gamma_++\Gamma_-)(\Gamma_++\Gamma_-+4\Gamma_z)}\right), 
\end{equation}
where we have again assumed $\cos(\phi)=1$ at resonance, allows the evaluation of the ratio $\Gamma_+/\Gamma_-$.

\section{Measurement of Coherent oscillations}
\label{sec:coherent}
A somewhat more demanding experiment would be to attempt to resolve coherent oscillations of the system.  This can be done with either Rabi oscillations \cite{Nakamura01}, or Ramsey oscillations \cite{Nakamura99,Hayashi03,Gorman05}.
Rabi oscillations are resonant driven oscillations of the type described in this paper, where the Rabi frequency is given by $E_{\rm ac} \gamma_x$.  In order to be able to observe Rabi oscillations, it is necessary that the oscillation period be long compared to the time resolution of the detector $t_{\rm det} < 1/(E_{\rm ac}\gamma_x)$, or alternatively, that the state can be shelved for measurement. In either case, Rabi frequency must be fast compared to the dephasing time $E_{\rm ac}\gamma_x > \Gamma_z+\Gamma_++\Gamma_-$.  We have calculated a distribution of Rabi frequencies, in units of the applied AC field strength, for a certain fabrication strategy that is described in the following section, and plotted the results in  Fig~.\ref{fig:random}d. From this we can see, that for an AC field strength of  $E_{\rm ac} = 0.01$MVm$^{-1}$, the value used for all calculations in this paper, and one which is possibly at the high end of what could be reasonable expected to be experimentally generated, the Rabi frequency is almost certain to be significantly less than $500 {\rm MHz}$.  This is likely to be slow compared to realistic dephasing rates, making resolution of Rabi oscillations unlikely.

For most quantum systems, and this case is no different, it may be more practical to attempt to measure the Ramsey oscillations.  These are the coherent oscillation between basis states that are not eigenstates of the system Hamiltonian.  The advantage observing Ramsey oscillation over Rabi oscillation is that the frequency of the Ramsey oscillations is significantly higher than the typical Rabi frequency. The frequency of the Ramsey fringes is given by the energy difference between eigenstates, $2 \Theta$, and can be increased with the application of a DC field $E_{\rm dc}$, to at least 40GHz. Apart from the obvious condition $2 \Theta > \Gamma_z+\Gamma_++\Gamma_-$, and assuming that the system can be shelved for measurement, the experimental requirements for resolution of these oscillations are that the measurement time is fast compared to the relaxation time $t_{\rm det} < 1/\Gamma_-$, and that the system can be manipulated on a timescale that is fast compared to the period of oscillations.

In General, Ramsey oscillations are significantly easier to observe than the Rabi oscillation, and have been observed in both superconducting systems \cite{Nakamura99,Vion03}, as well as coupled quantum dots \cite{Hayashi03}, whilst Rabi oscillations have only been resolved in superconducting systems \cite{Nakamura01,Vion03}.

\section{Uncertainties due to imprecise fabrication}
\label{sec:stats}
We now turn our attention to uncertainties that may arise due to the statistical nature of the fabrication procedure for such a device.  Atomically precise placement of phosphorus donors in beyond current technology, even the most reliable fabrication procedures lead to an uncertainty of at least several lattice sites in the final positions of the donors \cite{Schofield03}.  As always, there is a tradeoff here between precision and speed, and the most practical way to create such a device is via the  implantation of a pair of single  phosphorus ions \cite{Jamieson05}, which leads to significantly higher uncertainty in the donor positions.

\begin{figure}
\rotatebox{0}{\resizebox{9cm}{!}{\includegraphics{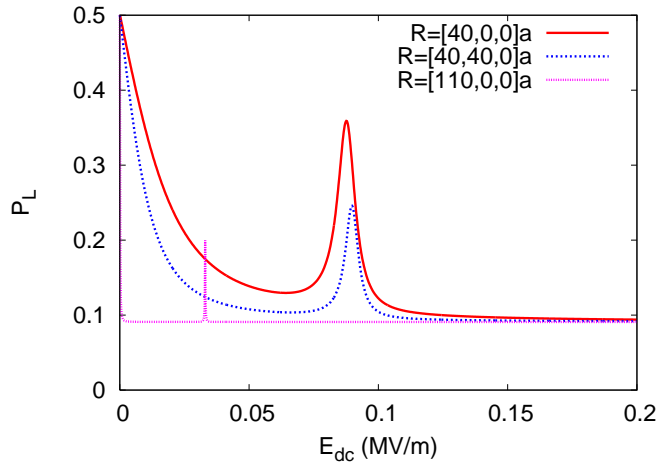}}}
\caption{(Colour online) Donor spectroscopy for several donor separations with significantly different separations.  This data plot is intended to illustrate that increasing the angle between the donor separation and the direction of the applied fields, has less effect on the spectroscopic system than increasing the donor separation.  All data were calculated with a $40$GHz driving field of amplitude ${\rm E_{ac}} = 10^{-2} {\rm MV/m} $.  The decoherence rates for the separations $R=[40a,0,0]$ and $R=[40a,40a,0]$  are $\Gamma_z = 3{\rm GHz}, \Gamma_- = 1{\rm GHz}$ and $\Gamma_+ = 100{\rm MHz}$, whilst for $R=[110a,0,0]$ we have used  $\Gamma_z = 300{\rm MHz}, \Gamma_- = 100{\rm MHz}$ and $\Gamma_+ = 10{\rm MHz}$, as the peak cannot be resolved with the above decoherence rates.}\label{fig:onetwo}
\end{figure}

Interactions of the implanted ion with the host silicon atoms gives rise to a phenomenon known as ion straggle, which results in an uncertainty in the final position of the ion.  Implantation of 14keV phosphorus ions in silicon results in a roughly Gaussian donor distribution, centred approximately 20nm below the silicon surface, and with a deviation $\sigma \approx 12nm$.  In addition, the relatively large spot size of the ion beam means that the ion is equally likely to enter the silicon substrate at any point in the mask aperture, which is approximately $30$nm in diameter.  The result is a complicated distribution of donor pairs, with a uniform distribution angles between the donor separation and the applied fields, decreasing the effective field strength.  An alternative method would be to implant a single ion into two separate apertures aligned with the field.  In Fig~.\ref{fig:onetwo} we show that the single aperture strategy is the best approach.  The Rabi frequency decreases as $\gamma_x ~ \cos(\theta) \exp(-R/a)$, where $\theta$ is the angle between the donor separation and the applied electric field, and so it is far more important to keep the donor separation to a minimum.

\begin{figure}
\centering
\begin{tabular}{cc}
a) \rotatebox{0}{\resizebox{8cm}{!}{\includegraphics{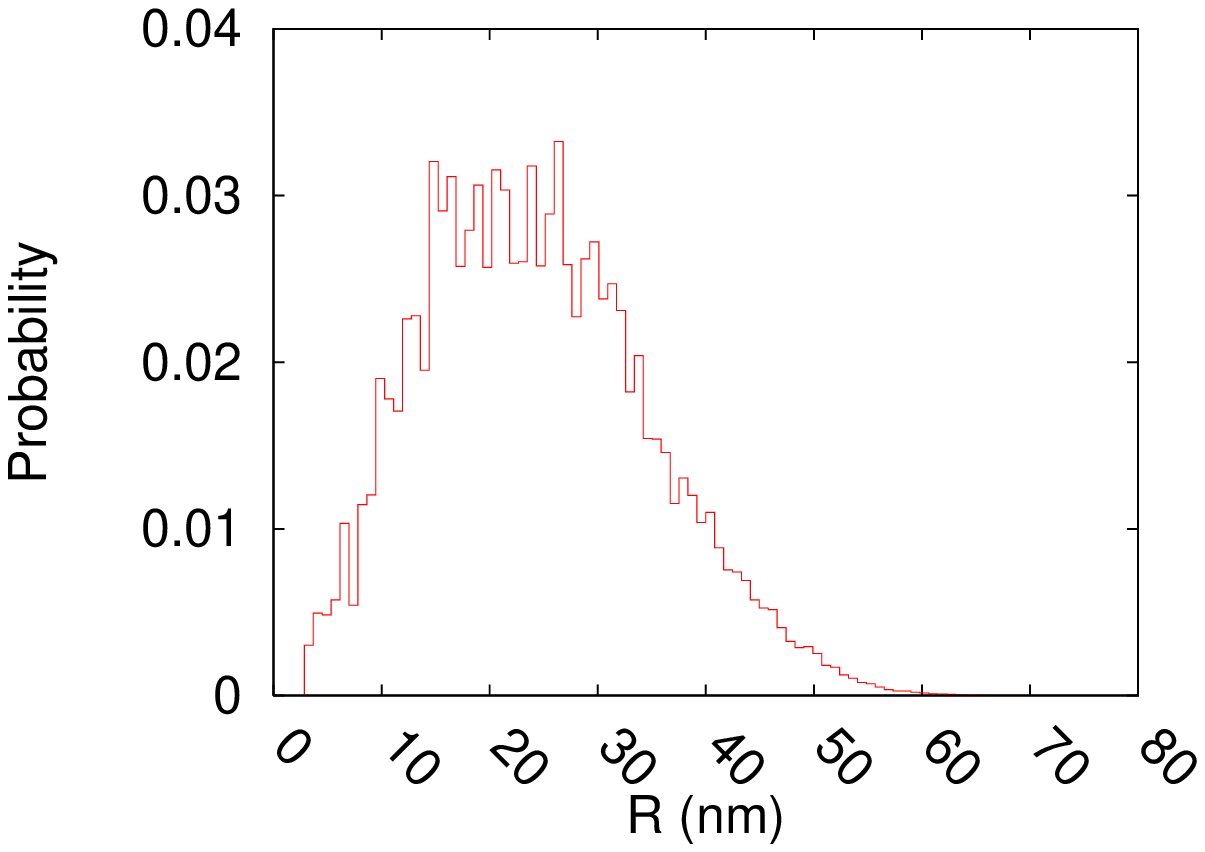}}} &
b) \rotatebox{0}{\resizebox{8cm}{!}{\includegraphics{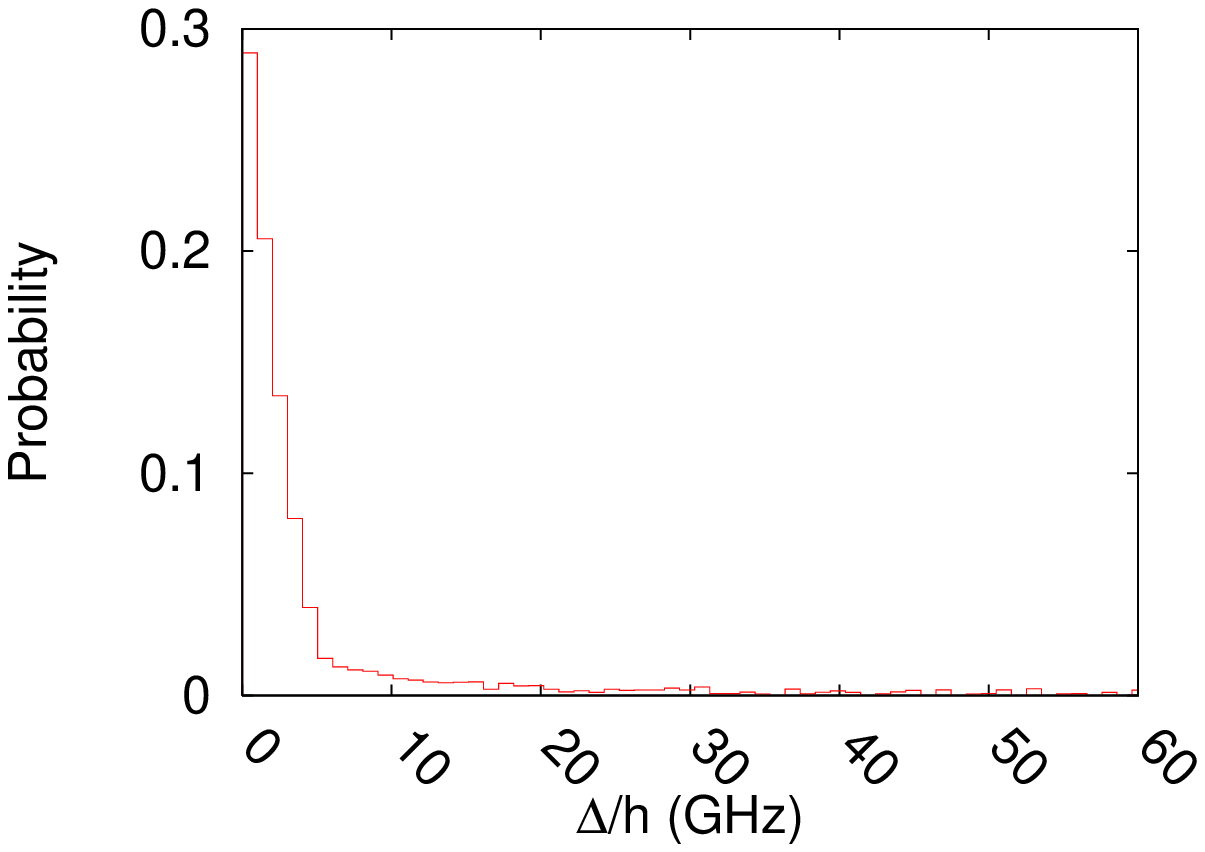}}} \\
c) \rotatebox{0}{\resizebox{8cm}{!}{\includegraphics{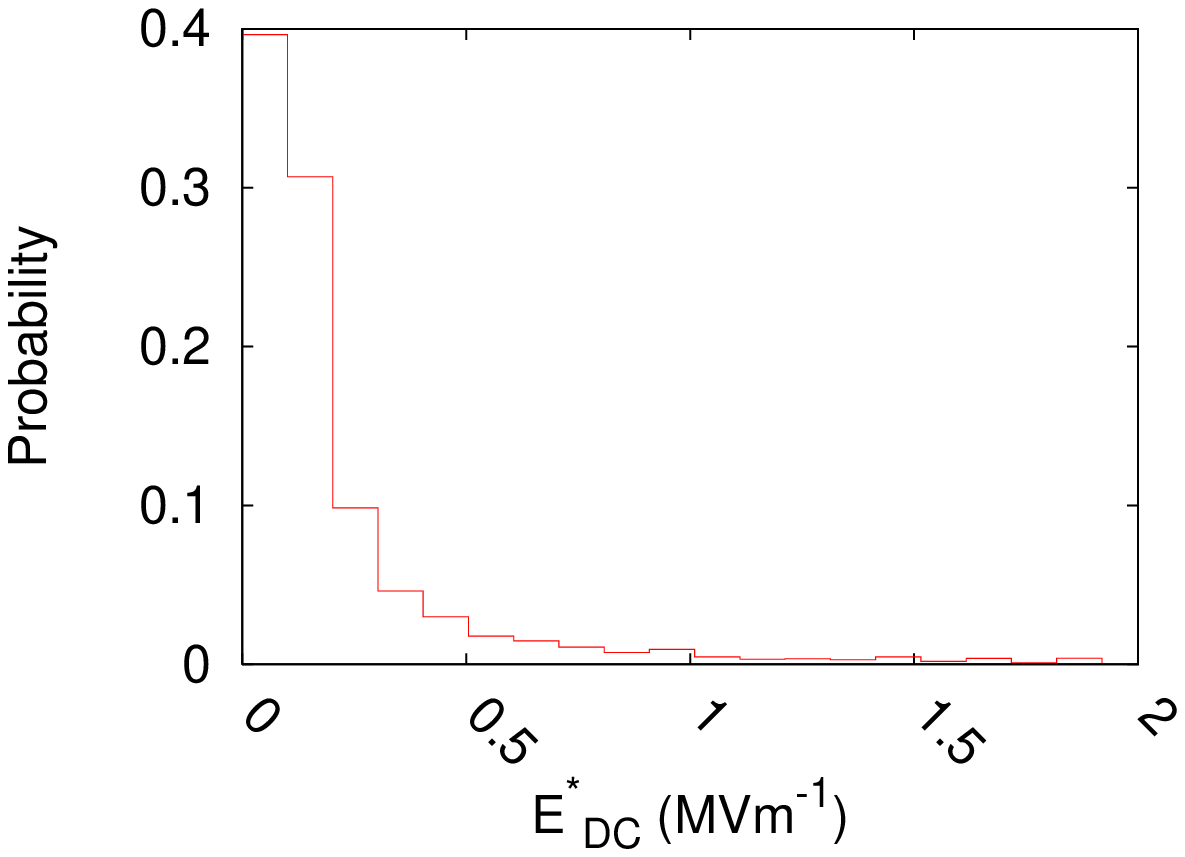}}} &
d) \rotatebox{0}{\resizebox{8cm}{!}{\includegraphics{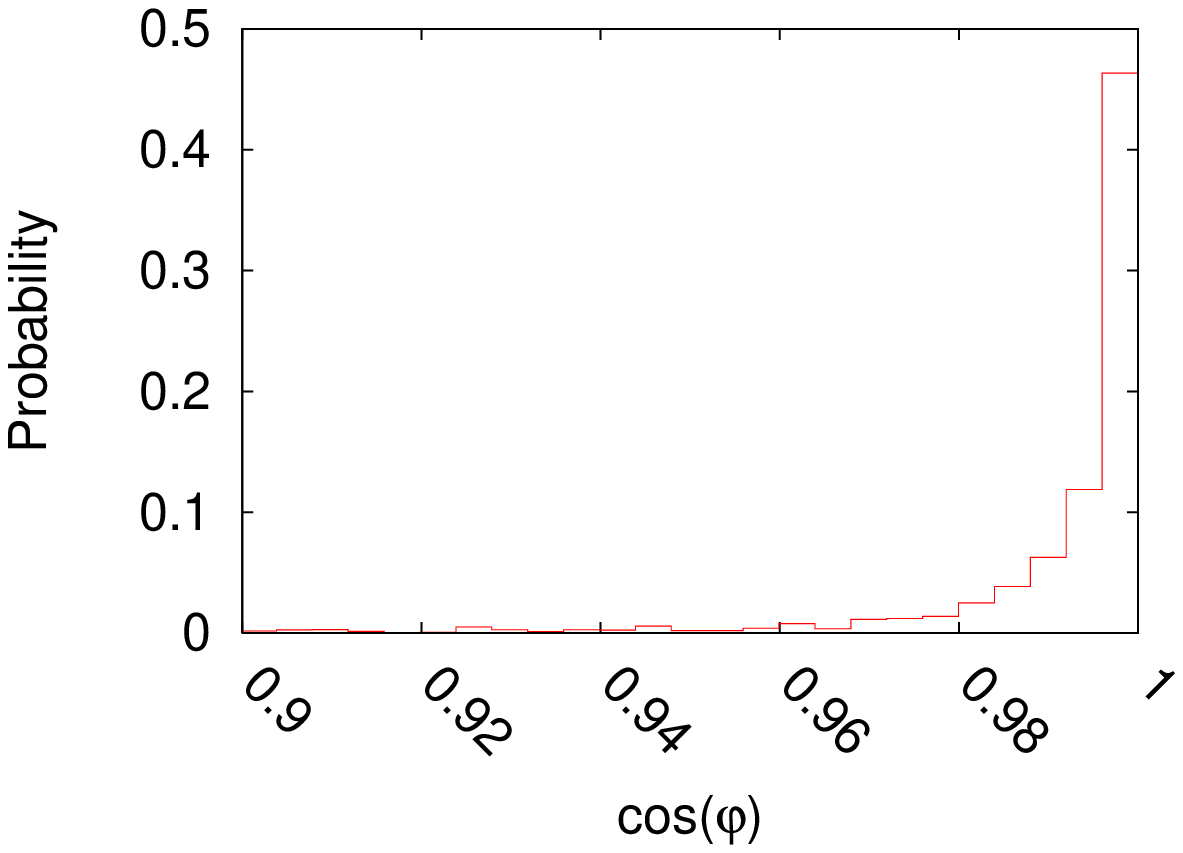}}} \\
e) \rotatebox{0}{\resizebox{8cm}{!}{\includegraphics{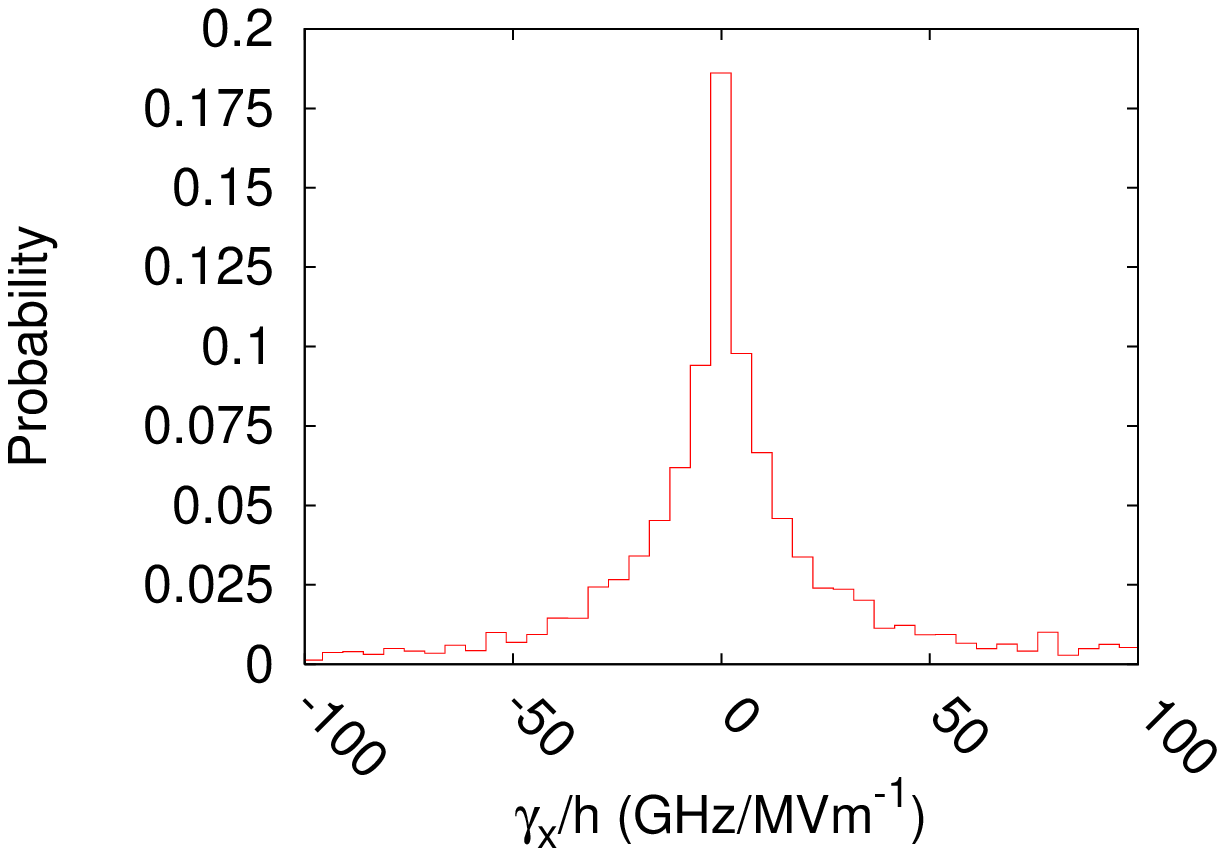}}} 
\end{tabular}
\caption{(Colour online) Distribution of system parameters for a 14keV implant strategy where the two donors are implanted 30nm apart along a 100 crystallographic axis, obtained using a total of 9149 points.  a) The distribution of donor separations R.  b) The distribution of the symmetric energy gap between the ground and first-excited states of the unperturbed qubit, $\Delta$.  c) The distribution of the DC field strength required to bring the system into resonance  with a $40GHz$ field.  d) The distribution of $\phi$, the angle that parameterises the degree of localisation of the ground-state.  e) The distribution of $\gamma_x$, the resonant Rabi frequency.}\label{fig:random}
\end{figure}

Using a typical distribution calculated for a 14keV ion implant in a single 30nm aperture, we have calculated the distribution of certain parameters of the system which affect the spectroscopic trace, using an ensemble of 16000 points.  These results are plotted in Fig~.\ref{fig:random}.  The distribution of the magnitude of the donor separation, is shown in Fig~.\ref{fig:random}(a), it is peaked around $R=30{\rm nm}$, and has a variance of the order of $10{\rm nm}$.  The distribution of the unperturbed energy gap $\Delta$  is plotted in Fig~.\ref{fig:random}(b), although it is strongly peaked at low frequency we see that there is a finite probability of finding $\Delta > 40$GHz.  In this case there is no chance of observing a single photon resonance peak using 40GHz microwave radiation, and although two-photon resonances may be observed, they will be of diminished magnitude.  We also note here that we have plotted the magnitude of the quantity $\Delta_{s-as}$ plotted in ref\cite{Hu05}, and so do not observe the negative values seen by these authors.  In Fig~.\ref{fig:random}(c) shows the distribution of $E^*_{dc}$, the DC field required to bring the system into resonance, which exhibits quite a large spread.  This is partly to do with the distribution of $\theta$, the angle between the donor separation and the applied electric field, clearly $E^*_{\rm dc} \propto 1/\cos(\theta)$, and so for large angles may be experimentally unachievable.  Fig~.\ref{fig:random}(d) shows the distribution of the quantity $\cos(\phi)$ at resonance.  This distribution is strongly peaked around 1, indicating that at resonance the eigenstates are likely to be the charge localised states, and the approximation made in deriving expression Eq~.\ref{eq:width} is likely to be valid.  Finally, Fig~.\ref{fig:random}(e) shows the distribution of Rabi frequencies, in units of the applied AC field strength, as discussed earlier, for achievable field strengths this is likely to be much lower than the frequency of Ramsey oscillations.

\section{Conslusion}

We have developed in this paper a detailed quantitative theory for the
electric field tuned microwave spectroscopic coherent response of the
electronic P$_2$ charge qubits in silicon.  Our realistic calculation, taking
into account fabrication and ion implantation aspects of the sample
preparation, shows convincingly that such an electric field tuned coherent
microwave response measurement to be feasible in the Si:P$_2$ quantum computer
architecture.  In addition, we establish that such a measurement would
directly provide information on the characteristic decoherence and
visibility times for Si:P charge qubits.  We find that the
"exchange-oscillation" type quantum interference problem \cite{Koiller02A},
arising from the six-valley degeneracy in the Si conduction band, does not
lead to any specific difficulties in the interpretation of the microwave
spectroscopy proposed herein.  Our proposed microwave experiment along with
the recent dc experiment proposed by Koiller {\it et al.} \cite{Koiller05}, which complements our proposal, are essential pre-requisites
for the eventual observation of Rabi and Ramsey coherent oscillations in the
Si quantum computer architecture.

\section{Acknowledgements}

C.W. and L.H.  acknowledge the financial support from the Australian Research Council, the Advanced Research and Development Activity (ARDA) and the Army Research Office (ARO) under contract number DAAD19-01-1-0653.  One of us (SDS) is supported by the LPS-NSA.  C.W. would also like to acknowledge discussions with Sean Barrett.

\bibliography{bibfile}% Produces the bibliography via BibTeX.

\end{document}